\documentclass{article}

\usepackage{arxiv}

\usepackage[utf8]{inputenc} 
\usepackage[T1]{fontenc}    
\usepackage{hyperref}       
\usepackage{url}            
\usepackage{booktabs}       
\usepackage{amsfonts}       
\usepackage{nicefrac}       
\usepackage{microtype}      
\usepackage{calc}
\usepackage{epsfig}
\usepackage{subcaption}
\usepackage{calc}
\usepackage{amssymb}
\usepackage{amstext}
\usepackage{amsmath}
\usepackage{amsthm}
\usepackage{multicol}
\usepackage{pslatex}
\usepackage{apalike,pifont}
\usepackage{graphicx}
\usepackage{doi}
\usepackage{cite}
\usepackage{amsfonts}
\usepackage{algorithmic}
\usepackage{graphicx}
\usepackage{textcomp}
\usepackage{multirow}
\usepackage{xcolor}
\usepackage{booktabs}
\usepackage{hyperref}
\usepackage{lscape}

\title{Up-to-date Threat Modelling for Soft Privacy on Smart Cars}

\author{ \href{https://orcid.org/0000-0002-7045-0213}{\includegraphics[scale=0.06]{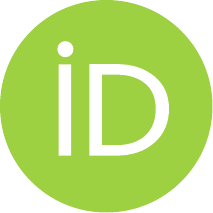}\hspace{1mm}Mario Raciti} \\
	IMT School for Advanced Studies Lucca\\
	Lucca, Italy \\
        Dipartimento di Matematica e Informatica\\
	Università di Catania\\
	Catania, Italy \\
	\texttt{mario.raciti@imtlucca.it} \\
	\And
	\href{https://orcid.org/0000-0002-7615-8643}{\includegraphics[scale=0.06]{orcid.pdf}\hspace{1mm}Giampaolo Bella} \\
	Dipartimento di Matematica e Informatica\\
	Università di Catania\\
	Catania, Italy \\
	\texttt{giamp@dmi.unict.it} \\
}

\renewcommand{\shorttitle}{Up-to-date Threat Modelling for Soft Privacy on Smart Cars}

\begin{document}
\maketitle

\begin{abstract}
Physical persons playing the role of car drivers consume data that is sourced from the Internet and, at the same time, themselves act as sources of relevant data. 
It follows that citizens' privacy is potentially at risk while they drive, hence the need to model privacy threats in this application domain.

This paper addresses the privacy threats by updating a recent threat-model\-ling methodology and by tailoring it specifically to the \textit{soft privacy} target property, which ensures citizens' full control on their personal data. The methodology now features the sources of documentation as an explicit variable that is to be considered. It is demonstrated by including a new version of the de-facto standard LINDDUN methodology as well as an additional source by ENISA which is found to be relevant to soft privacy. The main findings are a set of 23 domain-independent threats, 43 domain-specific assets and 525 domain-dependent threats for the target property in the automotive domain.
While these exceed their previous versions, their main value is to offer self-evident support to at least two arguments. One is that LINDDUN has evolved much the way our original methodology already advocated 
because a few of our previously suggested extensions 
are no longer outstanding.
The other one is that ENISA's treatment of privacy aboard smart cars should be extended considerably because our 525 threats fall in the same scope. 
\end{abstract}

\keywords{risk assessment \and automotive \and LINDDUN \and ENISA.}

\section{Introduction}
\label{sec:introduction}
Modern smart cars are full-fledged, interconnected nodes of a computerised ecosystem, often referred to as the Internet of Everything. The data that people generate while driving qualifies as personal data because it can be referred to drivers, and sometimes to their digital identity specifically. Therefore, it falls under the requirements of the General Data Protection Regulation in Europe, and of similar juridical prescriptions worldwide. 
The integration of various sensors, cameras, and communication systems in modern vehicles creates new opportunities for privacy breaches, raising concerns about data protection measures and corresponding risks. 

It follows that people’s privacy may be put at stake when they become car drivers. While \textit{hard privacy} concerns the various techniques to protect a subject's personal data from everyone else, such as anonymisation and minimisation, \textit{soft privacy} pertains to the range of practices to be followed for the subject to share their personal data with someone else while keeping full control, such as consent mechanisms and impact assessments.
Our research rests on the observation that privacy issues in the automotive domain are not fully understood at present, although they are certain to demand GDPR compliance. Compliance may be addressed in terms of privacy risk assessment, which in turn demands privacy threat modelling, hence the general motivation for this paper, 
which is spelled out more in detail in the sequel of this Section.

\subsection{Context and Motivation}
\label{subsec:context}

Privacy is a complex and multifaceted concept that may be interpreted in different ways in different contexts, yet we take it as a fundamental human right in the first place. In a GDPR fashion, we may summarise privacy as the right of an individual, that is, the data subject, to control or influence what information related to them may be collected, processed and stored, and by whom and to whom that information may be disclosed. Privacy and security are distinct concepts that should not be used interchangeably. While threat modelling has traditionally been approached from a security perspective, a challenge for all privacy threat modelling approaches comes from the following question: ``how to consider the impact on data subjects involved in the privacy threat?''
This aspect is stressed in law and regulation compliance, i.e., in the Data Protection Impact Assessment (DPIA), required under the GDPR, to help identify, assess, and mitigate privacy risks associated with data processing activities. Arguably, a DPIA would benefit from a privacy threat model.

Threat modelling is challenging as the analyst faces various problems, such as completeness and threat explosion. On the one hand, completeness may be impactful because failing to account for specific threats would cause pitfalls to the subsequent risk assessment.
On the other hand, the pursuit of completeness can result in a phenomenon known as threat explosion, characterised by an overwhelming number of threats that may be irrelevant, infeasible, or redundant with each other. Completeness and redundancy are considered by our previous work that features \textit{threat embracing}~\cite{vehits23}. Briefly, if two or more threats are described by labels that are deemed to be redundant in terms of their semantic similarity by the analyst's scrutiny, then these threats 
can be semantically merged into one.

Furthermore, as we shall see below, there is a lack of privacy threat taxonomies for smart cars in the state of the art, hence a clear motivation to push towards the advancement of a privacy threat modelling framework tailored for the automotive domain. Therefore, we modelled soft privacy threats for the automotive domain through a novel methodology that features a combinatoric approach~\cite{acsw23}. In short, we produced a final list of threats by taking a domain-dependent approach and by leveraging the threats from various sources, including in particular the LINDDUN state-of-the-art privacy threat modelling framework~\cite{landuyt2020} and ENISA's ``Good practices for security of smart cars''~\cite{enisa-report}. In particular, although the ENISA report is among the most relevant sources about car cybersecurity in Europe, its treatment of privacy is very limited, hence the need for a deeper close-up.

However, LINDDUN has recently been significantly updated, hence the results from our previous work demand an accurate revision. More precisely, LINDDUN has increased the number of soft privacy threats and,
in consequence, 
an up-to-date list of soft privacy threats for the automotive domain must be modelled. It could be pursued by leveraging the new version of the LINDDUN methodology, 
an approach that would bring the useful byproduct of checking how LINDDUN has evolved over time, particularly whether in the same direction we advocated~\cite{vehits23}.

\subsection{Research Question and Contributions}
Following the context and motivation given above, this paper focuses on soft privacy in the automotive domain from the threat modelling perspective. With the aim of advancing previous research, this paper addresses the core research question:
\begin{quote}
RQ1 \textit{What are the soft privacy threats for the automotive domain?}
\end{quote}

The following treatment answers the research questions by advancing an improvement of our innovative privacy threat modelling methodology~\cite{vehits23} and applying it to the current landscape of the automotive domain. A key advantage of our methodology lies in its combinatoric approach, which offers two key benefits: the elicitation of domain-independent threats by analysing relevant sources from the state of the art; the elicitation of domain-dependent threats by combining a generic threat knowledge base with domain-specific assets. Furthermore, by incorporating five variables into the analysis, our privacy threat modelling methodology ensures that the direction pursued by the analyst remains focused and aligned with the desired outcome. The variables act as guiding principles, allowing the analyst to make informed decisions based on relevant and reliable information.
The updated methodology adopts the mentioned ENISA report on smart cars as a source of specific and comprehensive knowledge on the automotive domain, and OWASP's ``Calculation of the complete Privacy Risks list 
v2.0''~\cite{owasp}.
In addition, these sources are augmented with the new version of LINDDUN
and with an additional representative of the state of the art, namely the ENISA ``Threat Taxonomy v2016''~\cite{enisa-threat-taxonomy}. Therefore, the new methodology rests on a significantly extended, domain-independent threat knowledge base. 

This paper contributes an updated version of our privacy threat modelling methodology and provides an updated list of 23 soft privacy threats that are domain-independent, thereby extending the 17 that we made available when we adopted the original LINDDUN~\cite{vehits23}. 
%
Because LINDDUN's soft privacy threats have increased from 9 to 17 over its two versions, our proposed extensions of it have decreased from 8 to 6. As we shall detail below, this can be taken as an indication that LINDDUN has evolved in the direction we advocated.

Moreover, our novel 23 threats are also combined with 43, rather than 41 as before, specific assets of the automotive domain, so as to produce, by appropriate combinations, a total of 525 domain-dependent soft privacy threats for the automotive domain --- each 
combination instantiates a given threat to each of the assets that are deemed affected by the threat.
These represent a substantial extension to the existing threat taxonomy introduced by the ENISA report on smart cars, which is rather scant in terms of privacy featuring only a couple of privacy threats. It could be argued that a better understanding of privacy within the automotive domain is achieved.

\subsection{Article Summary}
\label{sec:article-summary}
The organisation of the manuscript follows a simple waterfall style. Section~\ref{sec:related-work} outlines the related work, and Section~\ref{sec:background} gives an overview of LINDDUN and its latest changes. Section~\ref{sec:methodology} describes our novel privacy threat modelling methodology. Section~\ref{subsec:demo} demonstrates the methodology by applying it to the automotive domain along with a case study, and Section~\ref{sec:conclusions} concludes.

\section{Related Work}
\label{sec:related-work}

The challenges implicated by threat modelling led Wuyts et al.~\cite{8844639} to highlight the problems of current knowledge bases, such as limited semantics and lack of instantiating logic. Also, the authors discussed the requirements for a privacy threat knowledge base that streamlines threat elicitation efforts.

Furthermore, it is also noteworthy to recall that the process of threat modelling inherently implies assumptions and arbitrary decisions. Landuyt et al.~\cite{landuyt2020} highlighted the influence of assumptions to the outcomes of the analysis during the risk assessment process, more precisely in the threat modelling phase in the context of a LINDDUN privacy threat elicitation.

In addition, several attempts were made for the purposes of threat modelling in the automotive domain. Vasenev et al.~\cite{vehits19} were among the first to apply an extended version of STRIDE~\cite{stride} and LINDDUN~\cite{Deng2011} to conduct a threat analysis on security and privacy threats in the automotive domain. In particular, the case study is specific to long term support scenarios for over-the-air updates. Moreover, this work suggests that the privacy topic in the automotive domain has not reached the same level of maturity as cybersecurity.

In general, threat modelling is part of a wider process, that is risk assessment. Wang et al.~\cite{Wang2021} proposed a threat-oriented risk assessment framework tailored for the automotive domain, with the aim, among the others, of overcoming assumptions and subjectivity. This framework can be considered a precursor to ISO/IEC:21434\cite{iso21434}, which was defined a year later. Also, the authors applied STRIDE and the attack tree method for the threat modelling.

Moreover, Chah et al.~\cite{CHAH202236} applied the LINDDUN methodology to elicit and analyse privacy requirements of CAV system, while respecting the privacy properties set by the GDPR. Such attempt represents a solid baseline for the broader process of privacy risk assessment tailored for the automotive domain.
Finally, Bella et al.~\cite{Bella2023} advanced a dedicated risk assessment framework for privacy risks in modern cars. They proposed a double assessment, combining an asset-oriented ISO approach with a threat-oriented STRIDE approach.

The above works addressed crucial topics such as threat elicitation, threat knowledge base, privacy threat analysis and privacy risk assessment, both in general and specifically tailored to the automotive domain. However, to the best of our knowledge, there are no works advancing privacy threat modelling upon the basis of the de-facto standard LINDDUN, in its new version, with the aim of eliciting both domain-independent and domain-dependent soft privacy threats. These are the distinctive features of the present contribution.

\section{A Primer On (The New) LINDDUN}
\label{sec:background}
It is convenient to provide an introduction to LINDDUN before proceeding with the description of our methodology.
LINDDUN is a privacy threat modelling methodology, inspired by STRIDE, that supports analysts in the systematical elicitation and mitigation of privacy threats in software architectures. LINDDUN privacy knowledge base represents one of its main strengths, and it is structured according to the 7 privacy threat categories encapsulated within LINDDUN's acronym~\cite{Deng2011}. Recently, LINDDUN has been updated, and it is now available under three flavours from a lean to an in-depth approach: LINDDUN GO, LINDDUN PRO and LINDDUN MAESTRO. In particular, LINDDUN GO comes in the form of a card deck representing the most common privacy threats; LINDDUN PRO takes on a systematic and exhaustive approach, supported by the knowledge base; LINDDUN MAESTRO targets an enriched system description to enable more precise threat elicitation, yet it is still under development.

The first notable difference with the old version lies on the acronym, which puts more emphasis on the privacy threat types rather than on the privacy properties affected by threats. In fact, for the sake of comparison, the acronym that was previously expanded as
\textit{Linkability}, \textit{Identifiability}, \textit{Non-repudiation}, \textit{Detectability}, \textit{Disclosure of information}, \textit{Unawareness}, and \textit{Non-compliance}, has now been revised as follows:



\begin{itemize}
    \item\textit{Linking:} associating data items or user actions to learn more about an individual or group.
    \item\textit{Identifying:} learning the identity of an individual.
    \item\textit{Non-repudiation:} being able to attribute a claim to an individual.
    \item\textit{Detecting:} deducing the involvement of an individual through observation.
    \item\textit{Data Disclosure:} excessively collecting, storing, processing or sharing personal data.
    \item\textit{Unawareness \& Unintervenability:} insufficiently informing, involving or empowering individuals in the processing of personal data.
    \item\textit{Non-compliance:} deviating from security and data management best practices, standards and legislation.
\end{itemize}

The framework considers the state-of-the-art privacy threat types according to the privacy threat properties introduced by Pfitzmann~\cite{anon_terminology}. These are categorised as hard privacy and soft privacy properties. In particular, unlinkability, anonymity and pseudonymity, plausible deniability, undetectability and unobservability, and confidentiality (hiding data content, including access control) are under the umbrella of hard privacy; user content awareness (including feedback for user privacy awareness, data update and expire) together with policy and consent compliance are, on the other hand, soft privacy properties. 

LINDDUN provides a set of threats specific to privacy, named as ``threat catalogue'', in the form of threat trees. These privacy threat trees are inspired by the Secure Development Lifecycle (SDL)~\cite{sdlc} and reflect common attack patterns~\cite{linddun-nutshell} on the basis of state-of-the-art privacy developments, structured according to LINDDUN or STRIDE threat category and, in the previous version of LINDDUN, also to Data Flow Diagram (DFD) element type. In fact, the consideration of the DFD interactions has become more implicit in the new version of the framework, as the threat trees have become independent of the DFD element type, thus resulting in a significant diminution of the number of nodes as a side effect. The new guidance on how to link the Data Flow Diagram interactions rests now solely on the LINDDUN mapping table.

Threat trees provide a formal way to describe the security of systems based on a variety of attacks. Basically, the root node represents the ultimate goal, e.g., the threatening to a property, the children nodes embody different ways of achieving that goal, namely refinements, hence leaves represent basic-level attacks that can not be further refined. In addition, non-leaf nodes can be conjunctive (logic AND) or disjunctive (logic OR)~\cite{schneier1999attack}.

In the new version of LINDDUN, threat trees provide support to reason about applicability (criteria), factors that determine threat impact (impact), and examples of each characteristic pertaining to the threat (examples). The framework provides a different view of the threat trees in terms of detail, as it is possible to consult each tree at three different levels: Basic, Examples, All details.

\begin{figure}[ht]
    \centering
\includegraphics[width=\textwidth]{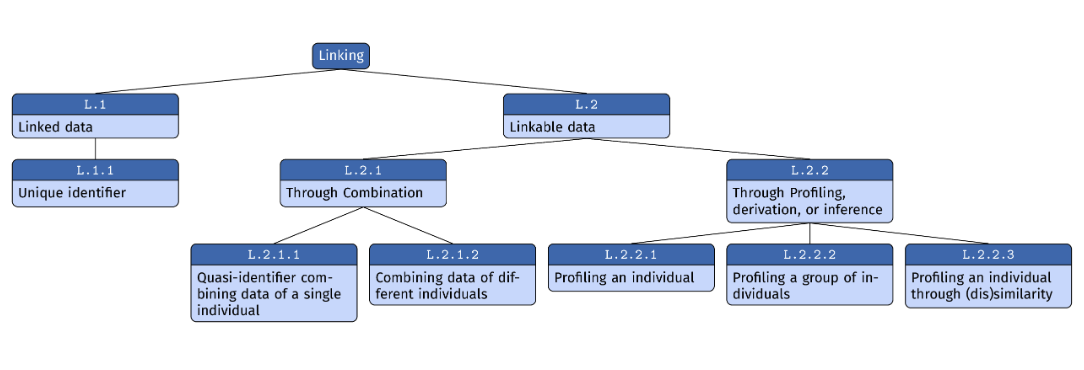}
    \caption{Example of a LINDDUN threat tree: Linking.}
    \label{fig:linddun-tree-example}
\end{figure}

An example tree is presented in Figure~\ref{fig:linddun-tree-example} for the Linking threat, which can be achieved through $L.1$ ``Linked data'', e.g., IP address, and $L.2$ ``Linkable data'', e.g., browser fingerprint. Both of these provide various attack paths which are not necessarily limited to the LINDDUN property analysed, namely Linking could lead to Identifying threats if we consider $L.1.1$ ``Unique identifier''.

We believe that the new version of LINDDUN represents a step forward from a GDPR perspective, as we can identify two LINDDUN privacy threat types, i.e., Unawareness \& Unintervenability (threats against data subject rights) and Non-compliance (violations against data protection principles), which tightly align with the European regulation by including as many as 17 threats. In the previous version of LINDDUN, these two types were already bound to soft privacy, but only included 9 threats. Moreover, these soft privacy threats were lacking relevant aspects, such as those related to data subject controls, consent, and violation of regulations, which are now caught by the new threat knowledge base. On the other hand, the remaining types target more technical privacy threats, gathered under the umbrella of hard privacy, and as such contribute more directly to the selection of ``appropriate technical and organisational protection measures''.

Despite LINDDUN threat trees may lack some formal semantics and have minimal selection criteria to express potential threats~\cite{8844639}, they still aim at providing a valuable overview of potential threat types that seeks to be general, hence suitable for a privacy threat analysis of any application domain. Moreover, the application of LINDDUN may lead to a high number of threats that may not be relevant, feasible, or important, thereby being labor-intensive and time-consuming, especially for complex or large systems. Hence, the advantage of having a catalogue of privacy threats, which are broad and applicable to various domains, may result in the problem of threat explosion.

\section{A Privacy Threat Modelling Methodology}
\label{sec:methodology}
This Section advances an improvement of our privacy threat modelling methodology~\cite{acsw23}. Our 
methodology incorporates both domain-independent and domain-specific knowledge and considers the potential consequences on the privacy of individuals as its cornerstone. The pivotal approach that our methodology relies upon is a combinatoric one with the aim of eliciting both domain-independent threats and domain-dependent threats. In particular, the former embody a generic threat knowledge base that consists of what is already known at present, whilst the domain-specific threats are derived from the first. Furthermore, in its previous version, our methodology identified four variables that contribute to model privacy threats, i.e., the specific privacy property, the threat agents, the application domain and the level of detail. 

\begin{figure}[ht]
    \centering
\includegraphics[width=\textwidth]{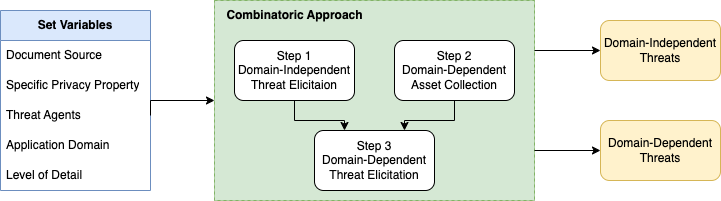}
    \caption{Diagram of our updated privacy threat modelling methodology.}
    \label{fig:diagram}
\end{figure}

The new version is enriched by considering an additional variable, that is, the document source.
The inclusion of five essential variables in our methodology orient the analysis, thus providing practical guidance to the analyst.
Figure~\ref{fig:diagram} depicts the updated methodology, while a description of the new introduced variable is provided below along with an outline of the combinatoric approach and a summary the other variables.

\paragraph{The Document Source}
In privacy threat modelling, the knowledge base is crucial for both threats and assets to be elicited. Threats and assets may be derived from different sources, e.g., state-of-the-art reports, scientific contributions, guidelines, et cetera. Therefore, the document source of the threats/assets that the methodology seeks to gather can be either \textit{internal} or \textit{external} to the analyst's institution. In the case of internal document source, threats/assets may arise from the analyst's expertise, knowledge of the particular institutional context, or insights into the specific system or domain being assessed. On the other hand, the external document source involves gathering threats/assets from external references, such as established best practices or recognised industry standards. This allows the analyst to leverage existing knowledge and insights from a broader community of experts.

A combination of both internal and external document sources may also be possible, for instance, when the analyst enucleates a new threat/asset being inspired from one or more external sources. In such a case, we refer to the document source of that threat/asset as \textit{hybrid}. Furthermore, the document source variable provides the means to keep track of the version of the threats, for example, the \textit{year} in which the specific threat list is published.
Moreover, when considering two or more different document sources, it may likely happen that some threats within such lists are inherently embraceable. Hence, threat embracing remains crucial for a proper merge of different document sources.

\paragraph{The Specific Privacy Property}
Privacy relates to the control that individuals have over their personal information, including how it is collected, used, and shared. According to the state of the art~\cite{danezis,Deng2011}, we can distinguish between two degrees of privacy, i.e., hard privacy and soft privacy. Briefly, while hard privacy focuses on minimising the risks associated with the collection and retention of personal data, soft privacy focuses on the appropriate use and sharing of personal data while respecting individuals' rights to control their data.
It is clear that, in addition to hard privacy and soft privacy, \textit{cybersecurity} plays a major, complementary role in terms of protection against the unauthorised access of data.

\paragraph{The Threat Agents}
The methodology refers to a threat agent as any entity, individual or group, who poses a threat to an individual's privacy. Unlike the security literature, which traditionally refers to such entities as ``adversaries'' or ``attackers'', here the term threat agent also includes other sources of risks for privacy, as a threat agent is less security-connotated and not limited to malicious actors only. In fact, we also consider three additional actors directly from GDPR, i.e., data controller, data processor, and third party as threat agents. 

\paragraph{The Application Domain}
The application domain in threat modelling identifies two prevailing approaches, i.e., domain-dependent and domain-independent. Domain-dependent threat modelling is specific to a particular application domain, such as healthcare, finance, or automotive, and it takes into account the unique characteristics of the domain itself, thus it may be more accurate and effective. On the other hand, domain-independent threat modelling is not specific to any application domain and can be applied to a wide range of systems. LINDDUN, for example, is a domain-independent methodology. A combination of the two approaches may offer a more effective and efficient analysis, picking the advantages of both.

\paragraph{The Level of Detail}
The level of detail of the statement describing a threat becomes relevant in the context of threat modelling and, subsequently, in risk assessment exercises with respect to the likelihood estimation of a threat. However, the most appropriate level of detail, that is, the choice of employing semantic relations, such as hypernyms or hyponyms, should be considered within the main picture, and the analyst will choose it with some inevitable bias.

\subsection{The Combinatoric Approach}
\label{subsec:combinatoric-approach}

The five variables introduced by our privacy threat modelling methodology are crucial in the execution of the combinatoric approach, as they contribute to follow the direction desired by the analyst.
The approach consists of three steps:

\begin{enumerate}
    \item Domain-Independent Threat Elicitation
    \item Domain-Dependent Asset Collection
    \item Domain-Dependent Threat Elicitation
\end{enumerate}

The first step involves the collection of domain-independent threats from relevant document sources.
The second step consists of the collection of a list of assets for the target domain from relevant document sources.

The third and last step aims at producing a list of domain-specific threats. In particular, for each domain-independent threat elicited in Step 1, this step associates to it the assets enumerated in Step 2. The sheer association expresses the object of the threat that was domain-independent in the first place, thereby making it domain-dependent. In other words, the domain-independent threat is instantiated over each of the assets it affects, producing a domain-dependent threat.

While relevant examples will be given in the next Section, if $\mathit{dit_1,\ldots,dit_n}$ is the list of domain-independent threats produced by Step 1, then
the number of domain-dependent threats that arise can be calculated as follows:
\[
\mathit{affected\_assets(dit_1)+\ldots+affected\_assets(dit_n)}.
\]


\section{Demonstration in the Automotive Domain}
\label{subsec:demo}
We apply our updated privacy threat modelling methodology to address the research question. In particular, we propose an exercise to focus on soft privacy with the new version of LINDDUN. While this paper details the key elements and findings, the full treatment is available online~\cite{repo}. The exercise is detailed below. In particular, we set the variables discussed through Section~\ref{sec:methodology} as follows:

\begin{itemize}
    \item[S:] External
    \item[P:] Soft Privacy
    \item[T:] Attacker, Data Controller/Processor, Third Party
    \item[D:] Domain-Dependent – Automotive
    \item[L:] Abstract
\end{itemize}

\subsection{Domain-Independent Threat Elicitation}
Soft privacy is the target property, therefore we must consider the LINDDUN threats that refer to such property, i.e., U(nawareness \& unintervenability) and N(on-compliance), as a first document source. For each node of the U-N property trees, we annotate the pertaining threat in a table.
It is convenient to provide a brief and general explanation of these threats, referring to the new descriptions provided by their sources.
In particular, U(nawareness \& unintervenability) refer to situations where individuals are not adequately informed, involved, or empowered in the processing of their personal data. N(on-compliance) refers to situations where a system deviates from security and data management best practices, standards, and legislation. It primarily focuses on the organisational and operational management context in which a system or service operates.



As one of the aims in eliciting the list of soft privacy threats is completeness, we may also want to extend the list of domain-independent threats by adding other \textit{external} document sources. In particular, our previous work~\cite{acsw23} included the 8 threats that were found~\cite{vehits23} to be outstanding with respect to the old version of LINDDUN. In detail, they account for the 2 threats from the ENISA report that fall under the ``Legal'' category, i.e., ``Failure to meet contractual requirements'' and ``Violation of rules and regulations/Breach of legislation/Abuse of personal data'', and the 6 threats from the ``Calculation of the complete Privacy Risks list 
v2.0''~\cite{owasp} document, i.e.,  ``Consent-related issues'', ``Inability of user to access and modify data'', ``Insufficient data breach response'', ``Misleading content'', ``Secondary use'', ``Sharing, transfer or processing through 3rd party''. 

These threats relate to soft privacy as per the definition of soft privacy that we covered previously in Section~\ref{sec:methodology}. Moreover, some of them are embraceable with the new threat catalogue proposed by LINDDUN. In particular, we notice that ``Violation of rules and regulations/Breach of legislation/Abuse of personal data'' is now \textit{embraceable} with several threats such as ``Regulatory non-compliance'' and ``GDPR''; ``Consent-related issues'' is now \textit{embraceable} with ``Invalid consent''; ``Inability of user to access and modify data'' with ``Lack of data subject control''; ``Insufficient data breach response'' with ``GDPR''. Hence, we can discard those threats, since they are already contemplated in the new LINDDUN threat trees, and keep the following ones: ``Failure to meet contractual requirements'', ``Misleading content'', ``Secondary use'', ``Sharing, transfer or processing through 3rd party''.

Moreover, we also consider here the ENISA ``Threat Taxonomy v2016''~\cite{enisa-threat-taxonomy} as another \textit{external} document source, as it is relevant to enrich the domain-independent threat knowledge base. We pick the threats that specifically target soft privacy. These can be found under the ``Legal'' category, i.e., ``Violation of laws or regulations/Breach of legislation'', ``Failure to meet contractual requirements'', ``Unauthorized use of IPR protected resources'', ``Abuse of personal data'', and ``Judiciary decisions/court orders''.
Again, three of such threats are already included in the more recent ENISA report on smart cars. In fact, ``Failure to meet contractual requirements'' is repeated and ``Violation of laws or regulations/Breach of legislation'' is embraced with ``Abuse of personal data'' into one single threat. Hence, we can add the following threats to the final list: ``Unauthorized use of IPR protected resources'', ``Judiciary decisions/court orders''.
It is noteworthy that these additions are still possible without consequences on the domain variable, as such threats are general privacy threats, i.e., they ignore domain specific entities. Hence, such threats can be analysed in relation with (virtually) any application domain.

In summary, we elicited a total of 23 soft privacy threats from the selected document sources, i.e., LINDDUN, ENISA and OWASP. Table~\ref{tab:new-soft} shows such threats --- the 6 that are highlighted are those that we do not deem embraceable with the current LINDDUN threats, hence represent our updated proposal for an extension to it. Moreover, while the 2 threats in italics are actually new (as they originate from the newly considered ENISA source), the remaining 4 already were among the 8 that we suggested before~\cite{vehits23}. It means that we managed to embrace half of the previous suggestions to current LINDDUN threats, something that we interpret as evidence that LINDDUN has been extended coherently with what we advocated.

\begin{table}[ht]
\caption{Domain-independent soft privacy threats elicited in Step 1.}\label{tab:new-soft} \centering
\begin{tabular}{|c|l|l|}
\hline
\multicolumn{1}{|l|}{{\textbf{S}}} & {\textbf{Threat}}     \\ \hline
\multirow{7}{*}{U}                     & Unawareness of processing                            \\ \cline{2-2} 
                                       & Unawareness as data subject                          \\ \cline{2-2} 
                                       & Unawareness as a user sharing personal data          \\ \cline{2-2} 
                                       & Lack of data subject control                         \\ \cline{2-2} 
                                       & Lack of data subject control – Preferences           \\ \cline{2-2} 
                                       & Lack of data subject control – Access                \\ \cline{2-2} 
                                       & Lack of data subject control – Rectification/erasure \\ \hline
\multirow{10}{*}{N}                    & Regulatory non-compliance                            \\ \cline{2-2} 
                                       & GDPR                                                 \\ \cline{2-2} 
                                       & Insufficient data subject controls                   \\ \cline{2-2} 
                                       & Violation of data minimization principle             \\ \cline{2-2} 
                                       & Unlawful processing of personal data                 \\ \cline{2-2} 
                                       & Invalid consent                                      \\ \cline{2-2} 
                                       & Lawfulness problems not related to consent           \\ \cline{2-2} 
                                       & Violation of storage limitation principle            \\ \cline{2-2} 
                                       & Improper personal data management                    \\ \cline{2-2} 
                                       & Insufficient cybersecurity risk management           \\ \hline\hline
\multirow{3}{*}{ENISA}                 & Failure to meet contractual requirements             \\ \cline{2-2}
                                       & \textit{Unauthorized use of IPR protected resources}          \\ \cline{2-2}
                                       & \textit{Judiciary decisions/court orders  }                   \\ \hline
\multirow{3}{*}{OWASP}                 & Misleading content                                   \\ \cline{2-2} 
                                       & Secondary use                                        \\ \cline{2-2} 
                                       & Sharing, transfer or processing through 3rd party    \\ \hline\hline
\end{tabular}
\end{table}

\subsection{Domain-Dependent Asset Collection}
For Step 2, we leverage two \textit{external} document sources from the state of the art, i.e., the assets identified in the work proposed by Bella et al.~\cite{Bella2023} and ENISA's taxonomy  of the key assets in the automotive domain. 
The former presents the following list of assets:

\begin{itemize}
    \item\textit{Personally Identifiable Information:} any data that could potentially be used to identify a particular individual, such as full name, date, and place of birth, driving licence number, phone number, mailing, and email address.
    \item\textit{Special categories of personal data:} data about the driver, e.g., racial or ethnic origin, political opinions, religious or philosophical beliefs, trade union membership, genetic data, biometric data, data concerning health or data concerning sex life or sexual orientation (GDPR art. 9).
    \item\textit{Driver’s behaviour:} driver’s driving style, e.g, the way the driver accelerates, speeds up, turns, brakes.
    \item\textit{User preferences:} data regarding cabin preferences, e.g., seating, music, windows, heating, ventilation and air conditioning (HVAC).
    \item\textit{Purchase information:} driver's financial information, such as credit card numbers and bank accounts.
    \item\textit{Smartphone data:} data that the vehicle and driver’s smartphone exchange with each other via the mobile application and short-range wireless connections such as Wi-Fi and Bluetooth (contact book, phone calls, text messages).
    \item\textit{GPS data:} vehicle geolocation history and route tracking.
    \item\textit{Vehicle information:} vehicle information such as carmaker, model, vehicle identification number (VIN), licence plate and registration.
    \item\textit{Vehicle maintenance data:} data about the maintenance and status of vehicle components such as kilometres travelled, tyre pressure, oil life, brake, suspension, and engine status.
    \item\textit{Vehicle sensor data:} data analysed and calculated by car sensors, such as distance sensors, crash sensors, biometric sensors, temperature sensors and internal and external cameras.
\end{itemize}

\begin{figure}[ht]
   \centering
\includegraphics[width=\textwidth]{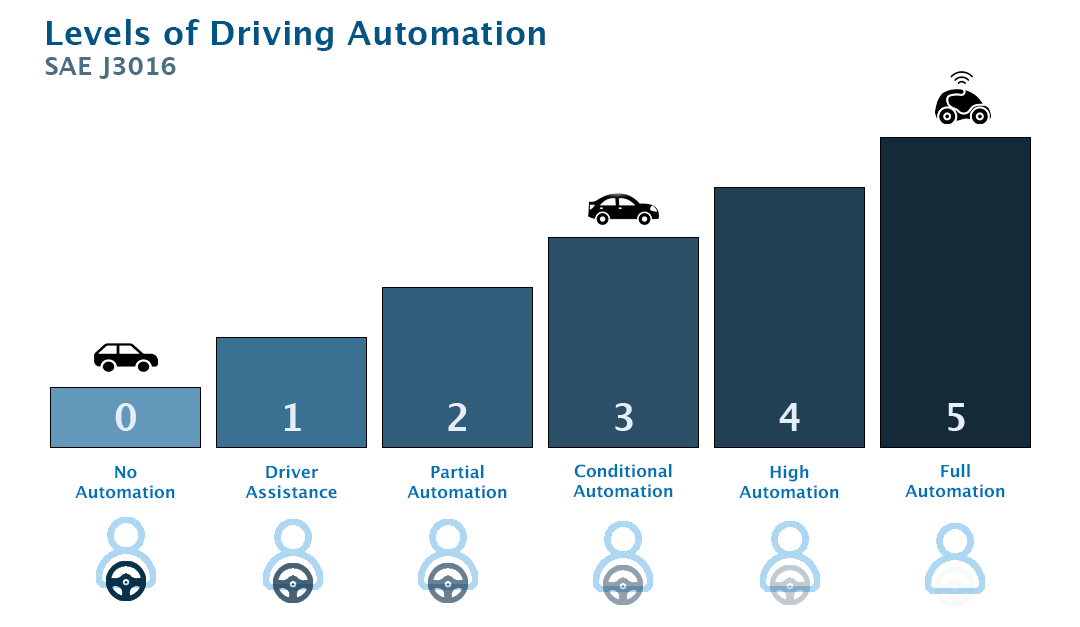}
   \caption{Vehicles automation levels as defined in SAE J3016.}
   \label{fig:sae}
\end{figure}

\noindent The report by ENISA focuses on Automated Driving System-Dedicated Vehicle (ADS-DS)~\cite{sae}, i.e., semi-autonomous and autonomous cars, and V2X communications, pertaining to SAE Level 4 and Level 5. Figure~\ref{fig:sae} depicts the SAE levels of driving automation.
The focus of the study is on smart cars that, as connected systems, have the necessary capabilities to autonomously perform all driving functions under certain (or all) conditions, and are able to communicate with their surroundings including other vehicles, pedestrians and Road-Side Units (RSU). Moreover, the key concepts analysed by ENISA do not only concern passenger cars but also commercial vehicles (e.g. buses, coaches, etc.), including self-driving, ride-sharing vehicles that can be shared with other users.

The assets proposed by ENISA are categorised in: ``Car sensors and actuators'', ``Decision Making Algorithms'', ``Vehicle Functions'', ``Software management'', ``Inside vehicle Communication Components'', ``Communication Networks and Protocols'', ``Nearby External Components'', ``Network and Domain Isolation Features'', Servers'', ``Systems and Cloud Computing'', ``Information'', ``Humans'', ``Mobile Devices''. For the sake of brevity, we only quote the descriptions of the assets under the ``Information'' category:

\begin{itemize}
    \item\textit{Sensors data} refers to data that is gathered by the
different smart car sensors and which will be transmitted to the appropriate ECU for processing.
    \item\textit{Keys and certificates} refers to the different keys and certificates used for security purposes (such as authentication, securing the exchanges, secure boot, etc.). Keys are stored in devices embedded in the vehicle (e.g. ECU) and/or in servers depending on their use.
    \item\textit{Map data} refers to the information about the car environment. Map data allows us to increase the passenger safety by correlating its information with the sensor data. Contrary to GNSS, which gives only information about the geolocalisation, map data gives information about the surrounding environment.
    \item\textit{V2X information} refers to the various information exchanged via V2X communications (e.g. emergency vehicle approaching, roadworks/collision warning and traffic information).
    \item\textit{Device information} refers to the various information related to a device embedded in a smart car (e.g. ECU, TCU) or connected devices (e.g. smartphones, tablet). This includes information such as type, configuration, firmware version, status, etc., of different smart car sensors and which will be transmitted to the appropriate ECU for processing.
    \item\textit{User information} refers to smart cars user (e.g. driver, passenger, etc.) information such as name, role, privileges and permissions.
\end{itemize}

Moreover, soft privacy is part of privacy, which is related to security, thereby all the assets proposed by ENISA may be virtually involved in the execution of the combinatoric approach.

During the execution of this step, within the list proposed by Bella et al., we identified some assets that are embraceable with the ENISA taxonomy. In particular, ``Personally Identifiable Information'' is \textit{embraceable} with ``User information''; ``Smartphone data'' with ``Device information''; ``GPS data'' with ``Map data''; ``Vehicle sensor data'' with ``Sensor data''. Thereby, we explicitly picked the following assets from the paper contribution: ``Special categories of personal data'', ``Driver's behaviour'', ``User preferences'', ``Purchase information'', ``Vehicle information', ``Vehicle maintenance data''. The last two were not available before. The remaining assets, according to our scrutiny, are already contemplated in the ENISA taxonomy.

Overall, we elicited a total of 43 assets, a small increase on the 41 that we had before~\cite{acsw23}.

\subsection{Domain-Dependent Threat Elicitation}
In the last step, we conjugate the findings from the previous steps. For each domain-independent threat elicited in Step 1, we assign the assets from Step 2 that we deem to be potentially affected by that particular threat. In general, a threat may apply to multiple assets, therefore for some threat-asset pairs we annotate multiple assets or, in case all assets are affected, we add the label ``All assets'' for the sake of brevity. In particular, most assets that we deem to be potentially affected by the soft privacy threats fall under the ENISA category ``Information''.

While the full results are available online~\cite{repo}, we present an exemplification of some noteworthy domain-dependent threats, with the additional aim of providing the rationale behind the related threat-asset(s) associations:

\begin{enumerate}
\item[$dit_{i_{1}}$] \textit{Unawareness of processing} refers to the lack of awareness or understanding about how personal data is being processed. It affects various assets, such as sensors data, map data, V2X information, device information, user information, special categories of personal data, user preferences, purchase information, vehicle information, and vehicle maintenance data.

\item[$dit_{i_{2}}$] \textit{Lack of data subject control – Preferences} specifically refers to the lack of control individuals have over their preferences. It affects assets such as user preferences and purchase information. When individuals cannot control or manage their preferences effectively, their privacy in relation to their preferences can be at risk. 

\item[$dit_{i_{3}}$] \textit{Regulatory non-compliance} encompasses all assets. It refers to the failure to comply with relevant privacy regulations or laws. When organizations do not adhere to the required privacy standards, all assets can be affected, leading to potential privacy breaches.

\item[$dit_{i_{4}}$] \textit{GDPR} is also associated with all assets. It specifically refers to non-compliance with the General Data Protection Regulation (GDPR), a data protection law in the European Union. Violations of GDPR can lead to severe penalties and legal consequences.

\item[$dit_{i_{5}}$] \textit{Violation of data minimization principle} refers to the violation of collecting and processing only the necessary data. It affects assets such as sensors data, map data, V2X information, device information, user information, special categories of personal data, user preferences, and purchase information, vehicle information, and vehicle maintenance data.

\item[$dit_{i_{6}}$] \textit{Unlawful processing of personal data} covers all assets. It occurs when personal data is processed unlawfully or without a legal basis. When personal data is processed in violation of applicable laws or regulations, it poses a significant privacy risk to all assets involved.

\item[$dit_{i_{7}}$] \textit{Lawfulness problems not related to consent} is associated with all assets. It highlights issues of lawfulness in data processing that are not specifically related to consent. These problems may include processing of personal data without a valid legal basis or exceeding the scope of permitted processing activities, such as automated decision-making on sensitive personal data.

\item[$dit_{i_{8}}$] \textit{Improper personal data management} is associated with user information and special categories of personal data. It signifies improper management practices regarding personal data, including inadequate safeguards, inappropriate handling, or unauthorised access. Improper data management can lead to privacy breaches, data leaks, or unauthorised use of sensitive information.

\item[$dit_{i_{9}}$] \textit{Failure to meet contractual requirements} refers to a breach of contractual requirements by Tier 1 and/or Tier 2 car components or software suppliers, thus encompassing all assets. Such threat may lead to financial, safety, privacy and/or operational impacts.

\setlength{\itemindent}{4mm}
\item[$dit_{i_{10}}$] \textit{Sharing, transfer or processing through 3rd party} refers to the sharing or transferring of various assets to third parties that increases the likelihood of unauthorised access, misuse, or breaches. It is clear that the affected assets belong to the ENISA ``Information'' category and include special categories of personal data, driver's behaviour, user preferences, purchase information, vehicle information, and vehicle maintenance data.

\end{enumerate}

As an outcome of this exemplification, the resulting number of domain-dependent threats would be:
\begin{tabbing}
\hspace{6mm}\=$\mathit{affected\_assets(dit_{i_1})+\ldots+affected\_assets(dit_{i_{10}})\ =}$\\[1.5ex]
\>$10+2+43+43+10+43+43+2+43+12=251$
\end{tabbing}

\subsection{Case Study}
\label{subsec:case-study}
This Section presents a case study that relies on the latest breaking news and articles about privacy incidents in the automotive domain. In particular, we employ classical web searches as a source of relevant information by building queries as ``privacy automotive'', ``automotive breach'', ``smart car privacy'', et similia, in the \textit{News} search filter offered by Google. If we matched some news with a soft privacy threat from the previous exercise, then we would be able to give some statistics about the occurrences of such threat, hence inferring an estimation of its likelihood. For the sake of brevity, we only present some illustrative examples of news that matched with one or more of the proposed soft privacy threats.
The following examples extend our previous case study~\cite{acsw23} and provide a different reading of the pieces of news in common, in light of the new threat list.

A data breach at Toyota Motor's Indian business~\cite{reuters} might have exposed some customers' personal information. ``Toyota Kirloskar Motor (TKM) has been notified by one of its service providers of an incident that might have exposed personal information of some of TKM’s customers on the internet''. This perfectly embodies a threat that we find in Table~\ref{tab:new-soft}, i.e., ``GDPR'', stemming from an inadequate response to a data breach that does not comply with GDPR.

Furthermore, we find another news that represents multiple threats: ``GDPR'', ``Lack of data subject control'', ``Insufficient data subject controls'' and ``Violation of data minimization principle''. The Dutch Data Protection Authority (DPA) investigated Tesla's camera-based ``Sentry Mode'' security system~\cite{automotive_news}, which is designed to protect the vehicle against theft or vandalism while it is parked. It does this by taking footage with four cameras on the outside of the vehicle.
This specific threat has now received a mitigation measure from the manufacturer, as the company altered security cameras to be more privacy-friendly and avoid GDPR violations. Originally, when Sentry Mode was enabled, this system was on by default. The cameras continuously filmed everything around a parked Tesla and stored one hour of footage each time.

In addition, we also found a review~\cite{cnn} that perfectly matched with the implications related to several soft privacy threats from the previous exercise. The article discusses a suggestion for a new feature to be added to the Ring Car Cam. The author proposes an Alexa-based voice command that would temporarily turn off the interior camera and microphone. This suggestion is based on the author's wife's volunteer work, which involves discussing private and privileged information about children's legal cases on the phone. The author's wife currently uses the physical privacy shutter to prevent the camera from recording video and audio inside the car. However, she sometimes forgets to flip the shutter up or down. Therefore, the author proposes a hands-free privacy trigger that would allow the user to enable or disable privacy mode with a voice command. This feature would eliminate the need for the user to physically manipulate the shutter, making it easier to maintain privacy while driving.

Moreover, we found a match for the ``Improper personal data management'' threat, as Toyota Japan~\cite{techcrunch} disclosed a significant data breach that occurred due to a cloud misconfiguration, resulting in the exposure of millions of customers' vehicle details over a decade. The exposed data included personal information, vehicle details, and videos.

Another discovery~\cite{bmw}, related at least to the ``Insufficient cybersecurity risk management'' threat, revealed that BMW may have potentially exposed sensitive files and client data, including customer information, as a result of an unprotected environment and the exposure of configuration files on the official BMW Italy website. Although the information alone may not compromise the website, it could be used for reconnaissance purposes by hackers. As a typical example of interconnection between privacy and security, the exposed configuration file could have allowed threat actors to find other vulnerabilities and access the site's source code.

The same interconnection between privacy and security is also tangible in the National Highway Traffic Safety Administration (NHTSA) warning~\cite{nhtsa} to carmakers in Massachusetts not to comply with a state law that requires them to share more vehicular telematics data with third parties. This naturally embodies the ``Judiciary decisions/court orders'' threat. The NHTSA argues that the state law is pre-empted by federal law and could potentially allow hackers to remotely access and control cars, leading to safety risks. The law, known as the ``right to repair'' law, has been the subject of a court battle between carmakers and the state. The NHTSA's letter represents the federal government's direct involvement in the case and raises concerns about the potential dangers of open access to vehicle telematics. The litigation is likely to face further delays due to the NHTSA's intervention.

\subsection{Evaluation}
\label{subsec:findings}

In this Section, we evaluate the findings from the previous experiment. The application of the combinatoric method to the automotive domain yielded notable results, which are available online~\cite{repo}, as stated above. In particular, we produced a novel, refined list of soft privacy threats that are domain-dependent. In fact, we associate the generic threat knowledge base pertaining to soft privacy, collected at the end of Step 1, with the automotive-specific assets collected at the end of Step 2, thus obtaining domain-specific soft privacy threats for modern cars with a homogeneous level of detail and dependent on the automotive domain, at the end of Step 3. 
A confirmation of the practicality and relevance of such threats for the automotive domain was proven by means of web searches.
This answers the research question.

Furthermore, the newly introduced variable in our privacy threat modelling methodology represents a foundational improvement for both Step 1 and Step 2, as the choice of the source document(s) requires a thorough examination.

In addition, it is important to emphasise that a crucial difference between the new list and the old list of threats was found: among the 8 threats added to the list in our previous work, 4 were deemed to be embraceable with the new LINDDUN threat catalogue. Hence, LINDDUN is clearly moving towards the direction that we hoped, and we are confident that their threat knowledge base will continuously improve in such a positive direction. Also, this supports the case that embracing is relevant and useful, especially when the analyst considers different document sources.

Our new list of threats enriches the broader threat knowledge base in the automotive domain over soft privacy. While we cannot claim that no more valid candidates exist, our final list of threats is complete with respect to the state-of-the-art knowledge base on soft privacy threats. Notably, such base features the new LINDDUN threat catalogue and the relevant taxonomies by ENISA. Our output is now available for the international community’s evaluation.

\section{Conclusions}
\label{sec:conclusions}
This paper faced the challenge of privacy threat modelling by focusing specifically on soft privacy and on the automotive domain.
Its research question found an answer through the development of an updated version of a previous threat modelling methodology, which now revolves around five rather than four variables.
These variables help the analyst make well-informed decisions upon the basis of a solid foundation of relevant and reliable data.

The methodology was demonstrated on a case study from the automotive domain, taking into account a new version of LINDDUN, which yields a de-facto privacy threat model, and an additional, relevant source by ENISA. As a result, as many as 23 domain-independent threats, 43 domain-specific assets and 525 domain-dependent threats for soft privacy in the automotive domain were produced~\cite{repo}. 

These results support the arguments that LINDDUN has evolved coherently with what we advocated before and that ENISA's privacy threats can be extended dramatically. Such arguments, in turn, represent a major leap forward in the modelling of soft privacy threats on smart cars.

\bibliographystyle{unsrtnat}
\bibliography{references}  






\end{document}